\documentclass[prb,twocolumn,showpacs]{revtex4}

\usepackage{graphicx}
\usepackage{amssymb}

\newcommand{\xplus}{$\mathrm{X}^{+}$}
\newcommand{\xmin}{$\mathrm{X}^{-}$}

\newcommand{\xminmin}{$\mathrm{X}^{2-}$}
\newcommand{\xx}{XX}
\newcommand{\x}{X}

\begin{document}

\title{Picosecond charge variation of quantum dots under pulsed
excitation}

\author{T. Kazimierczuk$^{1}$}
 \email{Tomasz.Kazimierczuk@fuw.edu.pl}
\author{M. Goryca$^{1,2}$}
\author{M. Koperski$^{1}$}
\author{A. Golnik$^{1}$}
\author{J.~A. Gaj$^{1}$}
\author{M.~Nawrocki$^{1}$}
\author{P. Wojnar$^{3}$}
\author{P. Kossacki$^{1,2}$}

\affiliation{
$^{1}$  Institute of Experimental Physics, University of Warsaw, Ho\.za
69, PL-00-681 Warszawa, Poland.\\
$^{2}$  Grenoble High Magnetic Field Laboratory, B.P.166, 38-042 Grenoble,
France.\\
$^{3}$  Institute of Physics, Polish Academy of Sciences, Al. Lotnik\'ow
32/64, 02-688 Warsaw, Poland  }

\date{\today}

\begin{abstract}

We present a spectroscopic study of excitation dynamics in self assembled
CdTe/ZnTe quantum dots. Insight into details of kinetics is obtained from
the time resolved  micro-photoluminescence, single photon correlation and
subpicosecond excitation correlation measurements done on single quantum
dots. It is shown that the pulsed excitation in energy above the energy gap
of the barrier material results in separate capture of electrons and holes.
The capture of carriers of different charge take place at different delay
from excitation.

\end{abstract}


\maketitle

\section{Introduction}

Quantum dots (QDs) belong to the most intensely studied topics in the
solid state physics. They owe their popularity to new physics involved and
to a wide range of their possible applications in such fields as
fabrication of efficient light sources, single photon emitters, information
storage and processing\cite{qd-fundamentals,Michler2000}. A particular
interest is related to the emerging field of quantum
information\cite{Loss-DiVincenzo,DiVincenzo}. 

Self-assembled semiconductor quantum dots receive an important share of
the research effort, due to efficient fabrication methods by modern
epitaxial growth techniques and possibilities of integration with existing
electronics.
QD studies started from the prototypical InAs/GaAs material system, and
were quickly extended over the entire families of III-V and II-VI
semiconductors\cite{Petroff2001,Bimberg-book}. 
Among the physical phenomena studied in the quantum dot research, those
related to light emission represent an important part, both under resonant
and non-resonant excitation.

All-spectroscopic methods are well suited to study excitation and light
emission processes in the QDs, in particular their
dynamics\cite{Bajracharya2007,Nee2006}.
The most precise information on the physical mechanisms involved in the
excitation and light emission processes is usually supplied by single QD
spectroscopy.

Basic QD spectroscopy methods include photoluminescence (both cw and
time-resolved) under varied experimental conditions: excitation power,
temperature, etc. In time-resolved studies the temporal resolution is
usually determined by the type of detectors used (down to tens of ps for
avalanche photo-diodes, to several picoseconds for streak cameras). 

In some cases, more sophisticated spectroscopic techniques are necessary.
For example, photon correlation measurements have been used to establish
that in the case of non-resonant excitation of QDs, carriers are trapped
separately rather than as whole excitons\cite{suff-prb_korelacje} 
(separate carrier capture in QDs was also demonstrated indirectly in cw
experiments \cite{Moskalenko2008, Matutano2008}).
Standard form of pump-probe techniques is not much used, as absorption
measurements of QDs present serious experimental difficulties
\cite{houel2007,alen2003}.

A technique more feasible for QD studies, similar to but not identical
with pump-probe methods, is excitation correlation spectroscopy
(ECS)\cite{rosen1981,olsson1982,Jorgensen83,ideshita1990,mishina1993,yamada1995}.
In ECS, photoluminescence is excited by pairs of laser pulses separated by
a
controlled delay. It has been shown to be a powerful tool to investigate
transient processes in semiconductors, especially excitonic recombination
\cite{Jorgensen83,Hirori06}. It has advantage of outstanding temporal
resolution limited only by the properties of light pulses. 
However, not all the possibilities offered by ECS have been exploited so
far. For instance, the order of carrier trapping in the excitation
processes has not been studied to the best of our knowledge.

In this work, we profit from the excellent temporal resolution of the
excitation correlation spectroscopy and apply it to a study of population
dynamics in single CdTe/ZnTe quantum dots (QDs). In particular we study
dynamics of carrier trapping by a QD under non-resonant pulsed excitation.
The choice of the CdTe/ZnTe system is motivated by its two advantages with
respect to the classical InAs/GaAs one. First, light emission in the
visible range (red to green), and second, more robust excitonic states
assuring efficient light emission at higher temperatures.

\section{Samples and experimental setup}

\begin{figure}
  \includegraphics{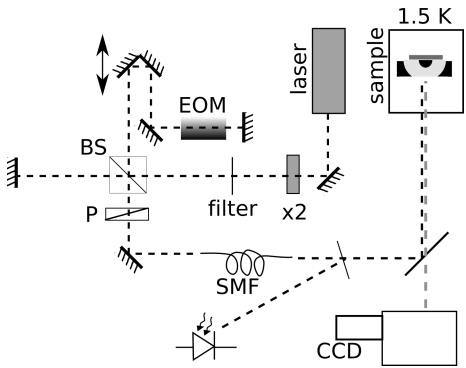}
  \caption{Experimental setup used in excitation correlation experiments.
The electro-optic modulator (EOM) was used to stabilize beam intensity
probed after passing by a~single mode fiber (SMF). BS denotes 50/50 beam
splitter and P linear polarizer. \label{fig1}}
  \end{figure}

The studied sample contained
an MBE-grown single layer of self-assembled
CdTe/ZnTe QDs. The sample growth was described in detail in Ref.
\onlinecite{wojnar08}. 
The density of quantum dots was estimated to be about $5\times 10^{9}$
cm$^{-2}$.
Measurements were performed on a sample immersed
in superfluid helium (at 1.8K). A reflection microscope, immersed
together with the sample, assured a spatial resolution better
than 0.5 $\mu$m. 
A frequency-doubled Sapphire:Ti femtosecond laser was used for pulsed
above-barrier 
excitation. 
In excitation correlation experiments, the sample was excited by pairs of
pulses with a controlled temporal separation (delay) between the pulses in
a pair. Consecutive
pairs were separated by the laser repetition period $13.6$ ns.
Time-integrated photoluminescence (PL) spectra were then recorded by a CCD
camera as a function of the delay. 
The experimental setup is presented
on Fig. \ref{fig1}.

Laser pulses were split in pairs in a Michelson interferometer setup. The
length of one arm of the interferometer was varied
by moving a~corner-cube retroreflector mounted on a~motorized translation
stage.
The setup allowed us to achieve a~controlled delay up to 4 ns. 
Beams from two arms of the interferometer were combined again on
a~50/50 beamsplitter forming a~train of pairs of pulses.
The joint beam was then transmitted
through 50 cm of single-mode optical fiber acting as
a spatial filter to assure a precise overlap of two laser
spots on the sample. The width of each~laser pulse 
at this point was estimated as 0.5 ps.
The most challenging task in such experiment was to assure the stability
of the
excitation of a single quantum dot. Due to imperfections in optical
alignment, 
the variation of the delay between the pulses
led to changes in the efficiency of coupling to the fiber. This effect was
canceled by introducing an~electro-optical modulator into the
variable-length arm 
of the interferometer, to stabilize the intensity of the laser after
the fiber. As a result, a good stability of excitation of a single quantum
dot was maintained over the measurement time which could exceed 6 hours.

In case of the single photon correlation measurements, a
Hanbury-Brown\&Twiss detection
scheme\cite{hbt_setup} was used. The photoluminescence from the sample was
split on a 50/50 beamsplitter and resolved by two monochromators equipped
with
avalanche photodiode (APD from Perkin Elmer or IdQuantique) single photon
detectors. 
The APDs were connected to
START/STOP inputs of TimeHarp 200 time counting system. An electrical
delay
introduced in the STOP signal allowed us to detect photons at negative
delay values.

\section{Photoluminescence spectrum of a~single QD}

\begin{figure}
\begin{center}
\includegraphics[width=85mm]{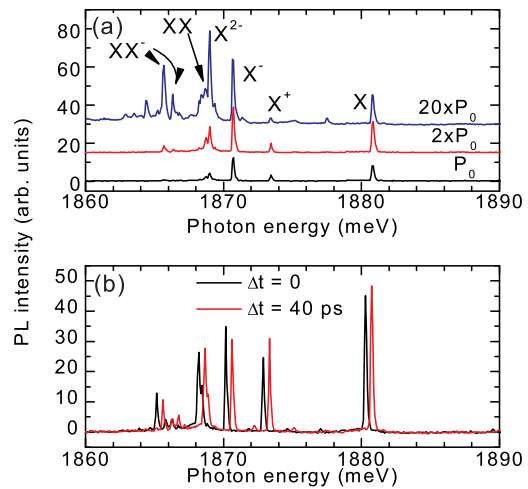}
\end{center}
\caption{(color online) (a) PL spectra of a~single QD under different
excitation intensities. (b) PL spectra excited by pair of laser pulses with
temporal separation $\Delta t$. The $\Delta t = 0$ps spectrum was shifted
horizontally for clarity by $-0.5$ meV.  \label{fig2}}
\end{figure}

Microphotoluminescence spectra of QD ensembles, limited by the size of the
excitation and detection spots, revealed an inhomogeneously broadened
distribution with a characteristic line structure.  A low density of the
lines in the low energy tail of the PL band allowed us to find well
isolated sets of lines originating from single quantum dots. An example PL
spectrum of a single quantum dot is presented on Figure~\ref{fig2}.
The lines were identified as originating from recombination of neutral and
charged excitons and biexcitons, 
as marked in the figure. The identification was based on relative emission
energies, in-plane anisotropy effects, 
and photon correlation measurements.

\begin{figure}
\begin{center}
\includegraphics[width=85mm]{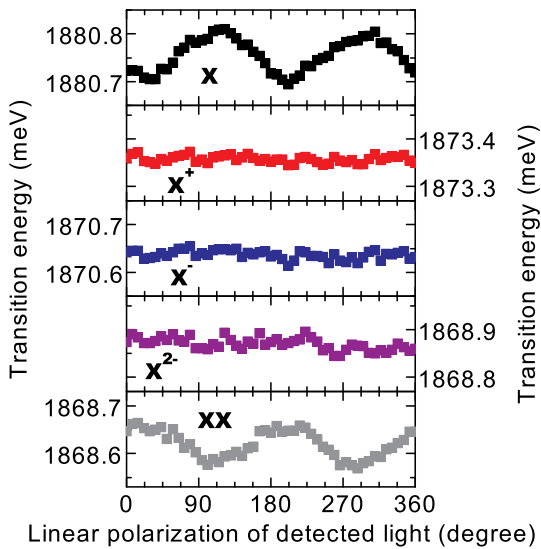}
\end{center}
\caption{(color online) Measured transition energy as a~function of
detected light polarization. Oscillatory behaviour indicates a small
anisotropic splitting. \label{fig3}}
\end{figure}

The lines emitted by the QD in the neutral or singly charged state were
first tentatively identified on the basis of the
characteristic pattern of their emission energies, observed in previous
experiments on similar samples\cite{kazimierczuk09}, in particular with
charge-tuning\cite{leger_1mn}. The identification of the lines related to
neutral exciton (X) and biexciton (XX) transitions was confirmed by
characteristic in-plane anisotropy effects. 
In case of a quantum dot of $\mathrm{C}_{2v}$ symmetry, both X and XX
lines are split in linearly polarized doublets, originating from the fine
structure splitting (FSS) of the excitonic state \cite{gammon_anisotropy}.
The experimental resolution did not allow us to observe the splitting
directly. However, the
components of the doublets could be observed in linear polarizations
parallel and perpendicular to the QD anisotropy axis. At intermediate
polarization angles, each doublet was observed as a single broadened line
at an intermediate spectral position.
This effect leads to oscillations of the apparent X and XX transition
energies as a function of orientation of detection polarization, as
presented in Fig. \ref{fig3}.
As expected, no energy oscillations were observed in case of charged
excitons, in particular \xplus and \xmin, which contain pairs of identical
carriers in singlet states.
An argument supporting the assignment of trion signs is a~negative optical
orientation of \xmin\ line, observed at quasi-resonant excitation through
a
neighbor dot \cite{kazimierczuk09}. The negative optical orientation had
been observed for negative trions in many QD
systems\cite{laurent-negative_orientation} and is related to electron-hole
flip-flop process.

Some linear polarization effects were also observed for the doubly charged
exciton \xminmin\ line. However, they are beyond the scope of this
work\cite{Kazimierczuk2009APP}.

\begin{figure}
\begin{center}
\includegraphics[width=8cm]{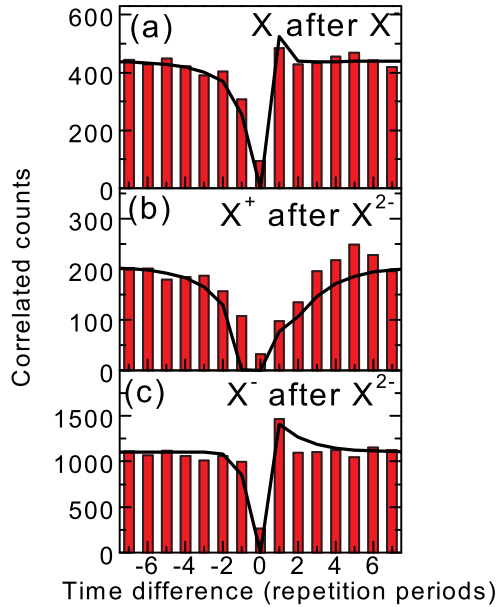}
\end{center}
\caption{(color online)Typical photon correlations related to change of
charge state. Negative time distance denotes pairs with inverted photon
order, e.g. \xmin\ after \x\ in case of (a). Solid lines were calculated
within a~model adapted from Ref. \onlinecite{suff-prb_korelacje} with
parameters $\alpha=0.80$, $\beta=0.86$, and $\xi=0.26$.
\label{fig4}}
\end{figure}

Further data supporting the identification of the lines were obtained from
photon correlation
measurements. In our experiments, the quantum dot was excited by
picosecond
pulses of light and photoluminescence photons related to selected
excitonic lines
were counted by two detectors. Example results of such experiments are
presented in Fig.
\ref{fig4}, in the form of histograms of detection events of pairs of
photons from the
two transitions as a~function of their temporal separation (number N of
laser repetition periods). Due to pulsed excitation, time delay between the
two emitted photons is close to integer multiples of the repetition period.
A clear antibunching (suppression of the peak) at zero delay confirms
unequivocally that we deal with a single photon emitter. Similar
antibunching was observed for autocorrelation experiments, whereas cascade
emission was witnessed by a characteristic bunching  (enhancement of the
central peak) in XX-X cross-correlation histograms (not shown).
The \xminmin\ line was identified using the cross-correlation histograms
presented in Fig. \ref{fig4}. Besides the central antibunching, they show
longer time-scale effects, extending over several repetition periods. Such
effects are known to originate from 
QD charge variation\cite{suff-prb_korelacje}. The relatively high
probability of the observation of correlated photons emitted after
adjacent
pulses in different charge states indicates an effective capture of single
carriers. In particular,
recombination of a neutral exciton after recombination of a negative trion
requires only a~single hole capture while a capture of three
carriers is necessary if the emission order is opposite. Therefore
corresponding probabilities of the events are respectively higher ($N=+1$
peak at Fig. \ref{fig4}a) and lower ($N=-1$
peak at Fig. \ref{fig4}a) than the probability in stationary state
($N\to\infty$). The similarity of \xmin\ - \xminmin\ and \x\ - \xmin\
correlation histograms supports the assignment of the line \xminmin\ 
to the doubly charged exciton. This assignment is confirmed
by relatively long characteristic time-scales of the \xplus\ - \xminmin\
correlation histogram, as it is related to the largest change of the QD
charge.

\begin{figure}
\begin{center}
\includegraphics[width=85mm]{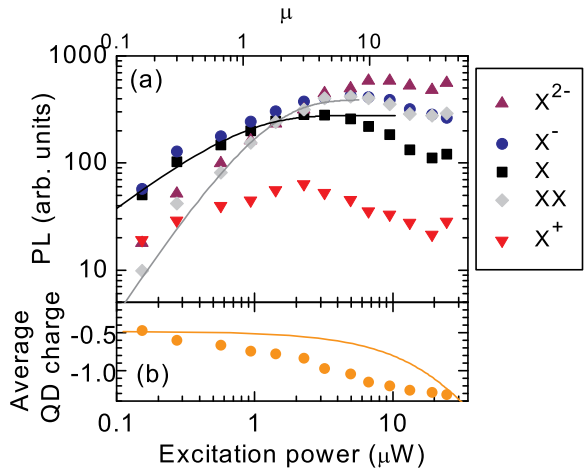}
\end{center}
\caption{(color online) (a) Photoluminescence intensity of various
excitonic lines and (b) estimated average QD charge state
versus excitation (pulsed laser) mean power. Solid~line was calculated
within a~model described in Section \ref{section:modelb} and shifted
vertically for clarity.
\label{fig5}}
\end{figure}

These qualitatively discussed correlation histograms can be simulated by
a~rate-equation model described in 
Ref. \onlinecite{suff-prb_korelacje}. We adapted this model by extending
the possible states to incorporate
transitions in $+1$ and $-2$ charge states. Free parameters of the model
include probabilities of capturing
an~electron, a~hole or a~free exciton. Solid lines in Fig. \ref{fig5} are
calculated with probability values taken from Ref.
\onlinecite{suff-prb_korelacje}. The results of the simulation confirm the
predominant role of the single carrier trapping. They depend weakly 
on exact value of free exciton capture probability. A similar model
is used in Section \ref{section:modelc}
to simulate time profile of excitation correlation results.

The simplest experiment which gives an insight into the excitation
dynamics is the measurement of photoluminescence spectra at various
intensities of the pulsed excitation.
A typical dependence of the intensity of selected excitonic lines on
excitation power is presented in Fig. \ref{fig5}(a).
In the lowest excitation limit, intensities of the PL lines exhibit
power-like dependence. In case of \x, \xplus, and \xmin\ transitions, the
dependence is linear. Two intensities (\xminmin and \xx ) increase
super-linearly. The superlinear dependence observed for the \xminmin line
is not surprising in view of a large number (four) of carriers necessary
to
form the \xminmin complex. Most of these results are in agreement with
typical behavior expected for single exciton complexes and for biexcitons
when the capture of whole excitons is significant
\cite{grundmann97,Abbarchi2009, suff-prb_korelacje}. 
A further increase of the excitation power leads to saturation of the line
intensities. This behavior is related to the fact that each excitation
pulse results at most in one recombination cascade (the excitation time is
much shorter than the recombination time). Therefore after one pump pulse
only one photon related to a determined transition may be emitted. Nearly
quadratic dependence of biexciton PL intensity indicates a~need for
including process of free exciton capture in addition to single carrier
capture evidenced by photon correlation experiments. 
One should note that even a relatively small exciton capture
rate may dominate over single carrier capture rates at low excitation
power in
multi-step excitation process, e.g. in case of XX formation.

At medium and high excitation power, the lines corresponding to negatively
charged states of the QD become relatively more intense. This effect can
be
analysed quantitatively in terms of the average charge of emitting QD
states calculated as:
\begin{equation}
\overline{q}_{\mathrm QD} = \frac{0\cdot I_\mathrm{X} + 1\cdot
I_\mathrm{X^+} - 1\cdot I_\mathrm{X^-} - 2\cdot I_\mathrm{X^{2-}}
}{I_\mathrm{X} + I_\mathrm{X^+} + I_\mathrm{X^-} + I_\mathrm{X^{2-}}}  
\label{eq:average_q}
\end{equation}
where $I_\mathrm{S}$ is PL intensity of line S. This formula approximates
the
averaged charge state of the quantum dot between excitation events since
only the fundamental transition of each observed charge state is taken
into
account. No matter which was the highest state in a recombination cascade,
its last step must be one of the final transitions: \x, \xplus, \xmin or
\xminmin.
Fig. \ref{fig5} presents the average QD charge state as a function of
excitation power. The effect of QD becoming negatively charged under
strong
excitation has been observed \cite{Moskalenko2008} but the underlying
mechanism cannot be determined without more detailed studies. It may be
caused by a modification of the electrostatic environment of the quantum
dot (similarly to that known for quantum wells \cite{Kossacki99,
Maslana}).
However, a mechanism inherent to the quantum dot itself is also possible.
It is related to the apparent absence of doubly positively charged states
of the quantum dot, while doubly charged negative trions have been
identified. 
The difference between number of bound states in positively and negatively
charged QD is predominantly caused
by the small valence band offset in the CdTe/ZnTe system.

\section{Measurements of excitation dynamics}

\begin{figure}
\begin{center}
\includegraphics[width=85mm]{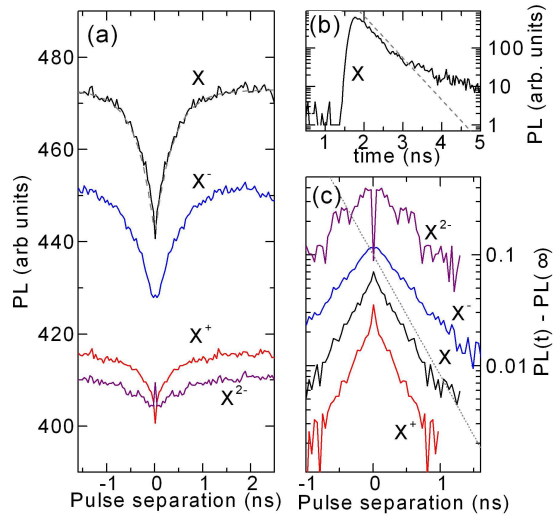}
\end{center}
\caption{(color online) (a) Excitation dynamics over long time-scale.
Presented data was symmetrized (averaged values for t and -t) for clarity.
(b) Photoluminescence decay after a~single laser pulse for neutral exciton
recombination. (c) Effect of two pulse excitation in a~semi-logarithmic
scale.
The dashed lines on all panels correspond to decay with 400 ps time
constant.
\label{fig6}}
\end{figure}

Excitation dynamics of the QDs was studied by means of excitation
correlation experiments.
As explained in Section II, a selected QD was excited by pairs of laser
pulses.
Time-integrated PL intensity was measured as a function of temporal
separation $\Delta t$
between the two pulses in the pair.
Plots of such dependence over the full temporal range are presented in
Fig. \ref{fig6}(a).

Features on two characteristic time-scales can be distinguished: a dip
several hundreds 
of ps wide, and a much sharper feature, both centered at zero delay. The
main effect 
is the relatively wide dip in the PL signal. Its width is comparable to
the radiative lifetime 
of excitonic states (Fig.  \ref{fig6}(b)). The effect arises near the
saturation regime 
when virtually each laser pulse excites the QD to a higher state. 
A qualitative explanation can be based on the fact that if the second
pulse in a~pair arrives
prior to the excitonic recombination then the second pulse does not
contribute to the intensity of the \x\ transition. On the contrary, when
pulses are separated by a few nanoseconds, they act independently
and therefore the recorded PL intensity is doubled with respect to single
pulse excitation. 
Thus, in the simplest approach the dip should be described by an
exponential function which may be directly compared to the decay of the
photoluminescence.
This is shown on Figure \ref{fig6}(b-c) where the dashed line presents a
profile obtained by fitting a monoexponential decay for $\left| \Delta t
\right| > 75\ \mathrm{ps}$ of the X profile from Fig.6a.

\begin{figure}
\includegraphics[width=85mm]{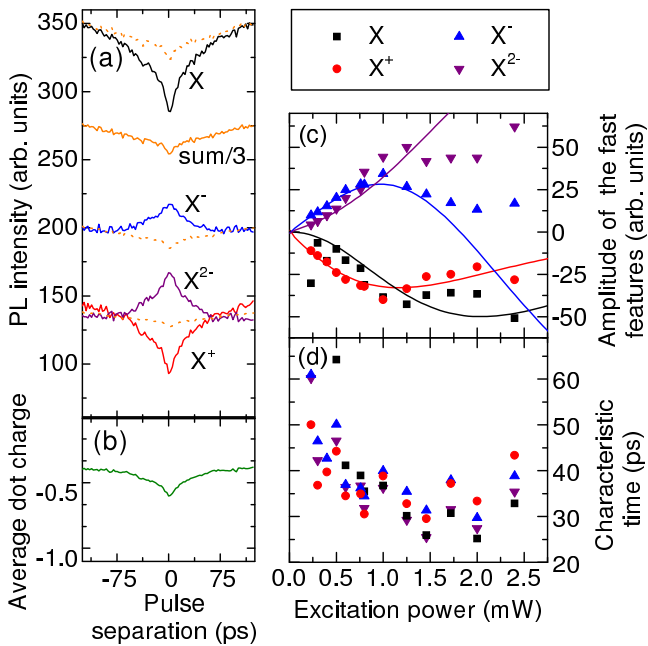}
\caption{(color online) Picosecond scale photoluminescence dynamics: (a)
example result of excitation correlation experiment for different excitonic
transitions and their sum (b) average charge state calculated according to
eq. (\ref{eq:average_q}). Power dependence of (c) amplitude and (d)
time-scale
obtained by fitting $a \left|\exp\left(-\Delta t / b\right)\right|$ to
baseline-corrected data (see text). Lines are calculated within a~model
described in Section V.B, with suitable x and y scale adjustments.
\label{fig7}}
\end{figure}

The second time-scale in the experiment is in the range of tens of
picoseconds. An additional variation of the PL intensity is observed within
this scale, as presented in Fig. \ref{fig7}(a). The signal is increased or
decreased depending on the excitonic complex with which the PL line is
related. A clear increase of the photoluminescence at zero delay is seen
for \xmin and \xminmin. A decrease is seen for neutral exciton and \xplus.
However, no significant effect is observed for a~sum $\Sigma_{\mathrm{PL}}$
of intensities of photoluminescence lines related to all the observed
charge states of the quantum dot (neutral and charged). This sum is shown
on Figure \ref{fig7} and its temporal variation is limited only to the slow
component related to the excitonic decay. 
Invariability of the sum suggests that the effect is related to the QD
charge state. Therefore we analyzed the averaged QD charge state versus
pulse separation. The calculated charge state is shown in Fig.
\ref{fig7}(b) and demonstrates a decrease of the averaged charge of the
quantum dot when the pulses are in coincidence. The characteristic time of
the feature appearing on the plots is the same as the lifetime of barrier
luminescence, observed on similar samples \cite{Korona2005}. Therefore we
might associate this feature to the process of the carrier capture by the
QD. The marked variation of the QD charge state during the pulse suggests a
possibility of non-synchronous capture of carriers of different charge. In
other words we wish to examine consequences of a delay between the arrival
of holes and electrons in the QD.

We parameterized the short-timescale feature using the following
procedure. In the first step we subtracted a~baseline originating from
long-timescale effect. We assumed for simplicity the same characteristic
long-timescale profile for each emission line and rendered it by a total PL
signal  $\Sigma_\mathrm{PL}$. We rescaled the total PL signal for each
emission line by a~constant factor to fit data points in a~range $100\
\mathrm{ps} < \left| \Delta t \right| < 125\ \mathrm{ps}$. 
The rescaled temporal profile was then subtracted from the profile of the
analyzed line.
In the second step, we fitted an empirical function $a \exp(-\left|\Delta
t / b\right|)$ to the refined data, obtaining an amplitude $a$ and a time
constant $b$ for each experimental scan.
The results of this procedure are presented in Fig. \ref{fig7}(c-d) versus
excitation power.

\section{Model description}

Experimental intensity-vs-delay plots, presented in the previous Section,
exhibit two main characteristic features, centered around zero delay: (i) a
sub-nanosecond decrease, common for most of the PL lines (with a
characteristic time comparable to the PL decay time), (ii) a feature on the
scale of tens of picoseconds, related to the variation of the QD charge
state.
A complete model description of the observed features is complex and
requires certain assumptions concerning excitation mechanisms of the
quantum dot, its relaxation channels and their characteristic times.
Therefore, to achieve a better insight in the physical mechanisms involved,
we first discuss simplified versions of the model, describing selected
characteristic features of the data. In part (A) we discuss sub-nanosecond
effects in the model neglecting details of the excitation process. In part
(B) we show how a delay between the capture of electrons and holes results
in a fast variation of the averaged charge of the quantum dot. The
amplitude of this variation is described in a simplified model, in which
both carrier capture profiles are completely separated in time, and only
their integrals are meaningful. In part (C) we include an analysis of the
temporal profiles and discuss characteristic times of the carrier capture. 

\subsection{Sub-nanosecond scale dynamics}
The shape of observed long-scale PL dependence can be explained by
introducing a simple analytical model.
Here we neglect the effects related to the QD charge state and consider
the QD energy spectrum as an infinite ladder of states, starting from the
lowest (ground) state for a given QD charge. Leaving out the charge degree
of freedom in this model is justified by a similar decay dynamics of all
the observed states (Fig.  \ref{fig6}(b)). 
Within this simple model, we assume that the number of captured e-h pairs
after a~single laser pulse does not depend on current QD state and is
described by a probability distribution $R(k)$. For example, in case of
free exciton trapping the probability of capturing exactly $k$ e-h pairs is
given by a Poisson distribution\cite{grundmann97,Santori2002,Abbarchi2009}
\begin{equation}
\mathcal{P}(k, \mu) = \frac{\mu^k}{k!}\mathrm{e}^{-\mu} \quad \mathrm{and}
\quad  \mathcal{P}(0,0) = 1
\end{equation}
where the average excitation $\mu$ is identified with the excitation
power. 
In case of separate capture of electrons and holes in a dot with limited
number of charge states, the probability distribution is close to the
square of the Poisson distribution. The particular shape of the
distribution does not change dramatically the sub-nanosecond effects.
After the pulse, the QD relaxes towards the ground state by radiative
decay.  We assume for simplicity that all the excited states have the same
lifetime of $\tau = 400 \mathrm{\ ps}$. 
Therefore the probability, that the quantum dot excited to $n^\mathrm{th}$
excited level emitted exactly $l$ photons over time $t$ is given by
truncated Poisson distribution:
\begin{equation}
\widetilde{\mathcal{P}}_n(l, t / \tau) = \left\{ 
\begin{array}{ll} \mathcal{P}(l, t/\tau) &  \textrm{if $l < n$} \\
\sum_{i=n}^\infty \mathcal{P}(i, t/\tau) & \textrm{if $l = n$} \\
0 & \textrm{if $l > n$}
\end{array} \right.
\end{equation}
Within this model, we derived the following expression for PL intensity of
the first excited state (i.e. exciton state) in an excitation correlation
experiment, when the two pulses are separated by $\Delta t$:
\begin{equation}
I(\Delta t) = \left(1 - R\left(0 \right) \right) 
\left(1 + \sum_{i=0}^\infty
R\left(i\right)\widetilde{\mathcal{P}}_i\left(i, \Delta t / \tau\right)
\right)
\end{equation}
Profiles of intensity versus $\Delta t$, simulated for different
excitation intensities $\mu$, taking $R\left(k\right) = \mathcal{P}\left(k,
\mu\right)$, are presented in Fig.~8.

\begin{figure}
\includegraphics[width=9cm]{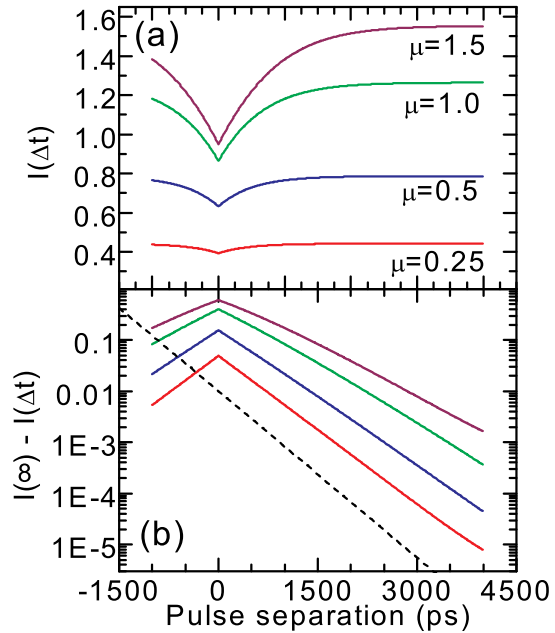}
\caption{(a) Simulated temporal profiles for different excitation
intensities $\mu$. (b) The same simulated data in scale proper for
a~relaxation processes. Function $\exp{(-t/ 400\mathrm{ps})}$
is shown for reference (dashed line).
\label{fig9}}
\end{figure}
It is interesting to note that the form of the simulated profiles is
approximately mono-exponential (Fig.~8b). However, their slope decreases
with increasing excitation intensity, and follows the lifetime of the
system only in the limit of low excitation intensity. This finding should
be kept in mind when interpreting the measurements in which correlation
excitation spectroscopy is used for determination of lifetimes
\cite{Jorgensen83, Hirori06}.
The stability of our experimental setup was not sufficient to verify the
predicted slope variation.
On the other hand, the assumption of equal lifetimes of all contributing
excited states is not fulfilled.

In spite of the simplicity of the approach, in which we neglect any
possibility of the change of the charge state of the quantum dot, the above
model reproduces quite well sub-nanosecond temporal profiles obtained in
the excitation correlation experiment. It gives also a quite good
prediction of the variation of the X and XX intensities versus excitation
power up to the saturation level. The predictions of the model for X and XX
lines are shown on Figure 5a by solid and dashed lines respectively. The
introduction of the possibility of the change of the charge state of the
quantum dot requires taking into account details of the carrier capture
after the excitation pulse.

\subsection{Picosecond scale dynamics \label{section:modelb}}
The understanding of the short-scale PL dynamics requires a different
approach. As mentioned previously, the effects on this scale are related to
the QD excitation process (mainly the QD charge variation) rather than to
the radiative relaxation after the first pulse. 

Generally speaking, the excitation correlation signal results from a
non-additive character (non-linearity) of the QD excitation by the two
pulses. This non-linearity may be attributed to processes occurring in the
barrier or in the QD.  Here we assume that both pulses generate the same
number of carriers in the vicinity of the QD, and the non-linearity
originates in the QD itself. To describe the observed QD charge state
variation, we assume that electrons and holes exhibit different trapping
dynamics after a~single excitation pulse. We will show that this approach
allows us to explain in a simple way all the observed experimental
features. In particular, we reproduce qualitatively the variation of the
averaged charge of the quantum dot with excitation parameters: excitation
power, delay between light pulses, and the ratio of energies of two
consecutive light pulses.

The non-synchronous trapping of holes and electrons may be caused by
different processes which result in different temporal profiles $g_e(t)$
and $g_h(t)$ of their capture rates. At this first stage, we will discuss a
simple overdrawn case. We consider the carrier capture profiles as
non-overlapping narrow pulses, with the electron trapping pulse delayed by
time $\tau_{e-h}$
with respect to the trapping of holes. The carrier trapping pulses can be
described by standard rate equations 

\begin{equation}
\begin{array}{lcl}
\dot{p}_+ & = & -g_e p_+  + g_h p_0                                       
                    \\
\dot{p}_0 & = &  g_e p_+  -\left(g_h + g_e\right) p_0  + g_h p_-          
                    \\
\dot{p}_- & = &            g_e p_0                     - \left(g_h +
\gamma g_e\right) p_-  +  g_h p_{2-} \\
\dot{p}_{2-}& = &                                         \gamma g_e p_-  
                 -  g_h p_{2-} 
\end{array}
\end{equation}
where we introduced an additional parameter $\gamma$ to account for
electron-electron blocking. The value of $\gamma$ was set at $0.3$ on the
basis of relative intensities of neutral-to-charged exciton line. Under our
assumptions, the rate equations can be integrated separately for the
electron and hole capture pulses, producing matrices $\mathbb{A}$ and
$\mathbb{B}$, describing the influence of the pulses on the charge state
probabilities.

\begin{equation}
\mathbb{A} = \exp{\left[
\begin{array}{cccc}
0 & \Gamma & 0 & 0 \\
0 & -\Gamma & \Gamma & 0 \\
0 & 0 & -\Gamma & \Gamma \\
0 & 0 & 0 & -\Gamma
\end{array}
 \right]}, \ \ \ 
\mathbb{B} = \exp{\left[
\begin{array}{cccc}
-\Gamma & 0 & 0 & 0 \\
\Gamma & -\Gamma & 0 & 0 \\
0 & \Gamma & -\gamma \Gamma & 0 \\
0 & 0 & \gamma \Gamma & 0
\end{array}
\right]}
\end{equation}
where $\Gamma$ denotes the integral of capture rate over the pulse,
assumed to be equal for electrons and holes and to be proportional to the
intensity of the laser beam. 

Each pair of laser pulses in the excitation correlation experiment will
produce two pairs of carrier capture pulses. Depending on the separation
between the two laser pulses, the hole trapping after the second pulse will
take place before or after trapping of electrons from the first pulse.
Thus, there are two possible orders of carrier trapping:
hole-hole-electron-electron or hole-electron-hole-electron. 
 The two cases can be described in terms of a recursive equation that
binds charge state distributions before and after a pair of excitation
pulses. For example, for the hole-electron-hole-electron ordering:
\begin{equation}
\left[ \begin{array}{c} p^{\mathrm{(after)}}_{+} \\ p^{\mathrm{(after)}}_0
\\ p^{\mathrm{(after)}}_{-} \\ p^{\mathrm{(after)}}_{2-} \end{array}
\right]
= \mathbb{A} \cdot \mathbb{B} \cdot \mathbb{A} \cdot \mathbb{B} \cdot
\left[ \begin{array}{c} p^{\mathrm{(before)}}_{+} \\
p^{\mathrm{(before)}}_0 \\ p^{\mathrm{(before)}}_{-} \\
p^{\mathrm{(before)}}_{2-} \end{array} \right]
\end{equation}
We computed stationary states in both cases and used the difference
between them as a measure of the amplitude of the picosecond-scale feature.
The results of the simulation are compared with the experimental results in
Fig. \ref{fig7}(c), after an appropriate adjustment of both amplitude and
power scales. A good agreement is achieved at low excitation power, while
some discrepancies appear at higher power. They may originate from the
absence of higher charge states in the model description.
The model provides also a correct qualitative description of the observed
evolution of the average QD charge state towards more negative values under
increasing excitation power, as shown in Fig. \ref{fig5}(b).
The enhancement of the negative QD states is obtained only if $\tau_{e-h}
> 0$, that is if electrons are captured after holes. 

\begin{figure}
\includegraphics[width=8.5cm]{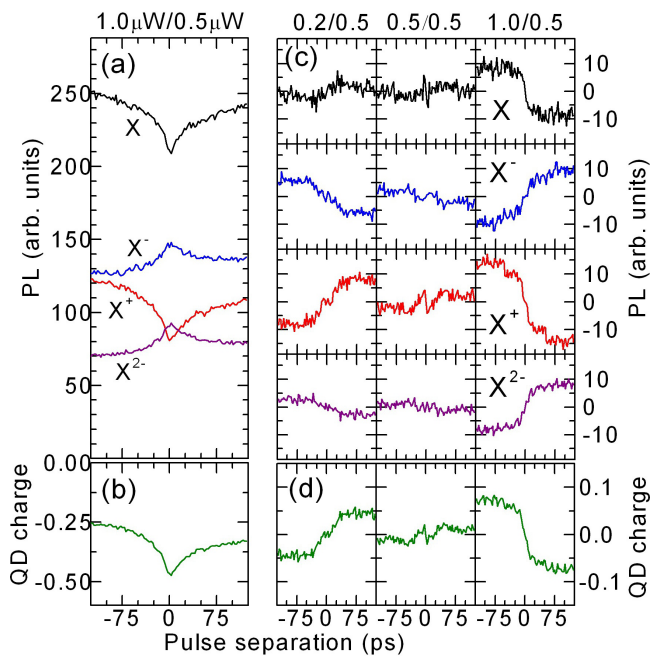}
\caption{(color online) (a,b) Results of excitation correlation experiment
with two pulses of different intensities (0.5$\mu$W/1.0$\mu$W denotes pulse
order for negative pulse separation). (c,d) Plots of effect asymmetry
obtained by subtraction: $PL(t) - PL(-t)$
for three sets of pulse intensities: 0.2/0.5$\mu$W, 0.5/0.5$\mu$W, and
1.0/0.5$\mu$W.
\label{fig9}}
\end{figure}

\begin{figure}
\includegraphics[width=8.5cm]{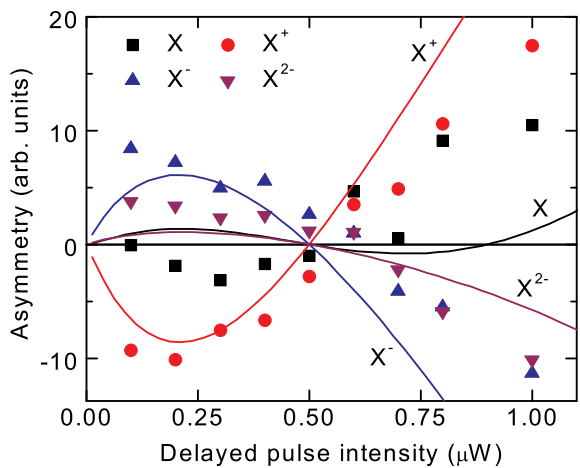}
\caption{(color online) Amplitude of asymmetry plotted against intensity
of one of the pulses. Intensity of the other pulse was set to 0.5 $\mu$W.
Solid lines are calculated within a~model described in Section V.B and
rescaled by a~common factor.
\label{fig10}}
\end{figure}

An additional test of the model was provided by experiments in which the
intensities of two consecutive light pulses were different. This eliminates
the temporal symmetry of the obtained profiles. In our experiments, the
power of one beam was kept constant and the power of the second one was set
at different levels in a series of consecutive measurements. Typical
experimental profiles, obtained for the power ratio 1:2, are shown on Fig.
\ref{fig9}(a). The obtained profiles are clearly asymmetric. This asymmetry
is better visible in the plots of difference between signals measured at
opposite delays. An example is presented in Fig. \ref{fig9}(c) for
different power ratio values. At ratio 1:1, as expected, the signal is
almost zero for all the excitonic complexes. For ratio different than 1,
the lines related to negative exciton complexes are stronger when the
stronger pulse arrives second. At the same time, neutral and positively
charged exciton lines are less intense. The signal asymmetry increases
during the first tens of picoseconds, reaches a maximum for a delay of
about 100ps and then decays with a decay time similar to the exciton
recombination time. We extracted the asymmetry amplitude and compared it to
the predictions of the model. The amplitude obtained for different exciton
complexes and ratios of the light power in two beams is presented on Fig.
\ref{fig10} and marked by symbols. In the model, different powers of the
two beams were simulated by taking different values of $\Gamma$
(proportional to beam power) in two pairs of matrices $\mathbb{A}$ and
$\mathbb{B}$ in equation 7. The results of the simulation are presented by
lines on Figure \ref{fig10}. The agreement is quite good and supports our
interpretation of separate capture of electrons and holes.

\subsection{Continuous rate equation model \label{section:modelc}}

The simple model discussed in section B was sufficient to analyze the
amplitudes of the picosecond scale features observed in the excitation
correlation experiment. However, to describe the temporal shape of the
observed peaks, we need certain assumptions about the profiles of the
carrier capture rates. Our data do not allow us to determine exact
profiles, nevertheless they give some insight in the characteristic times.
We propose here a rate equation model with simple exponential decays of the
hole and electron capture rates $g_h(t)$ and $g_e(t)$. They both start at
the time of arrival of the light pulse and decay with different time
constant $\tau_h$ and $\tau_e$ respectively. Such profiles could be related
to the exponential decay of free carriers in the barrier material and/or
wetting layer. The free carriers could be trapped by quantum dots or other
centers. Direct measurements of the time resolved photoluminescence from
the barriers in similar samples show a fast monoexponential decay
\cite{Korona2005}. Such decay would be a straightforward consequence of the
above assumption if $\tau_e \gg \tau_h$, and then the measured PL decay was
equal $\tau_h$.

\begin{figure}
\includegraphics[width=9cm]{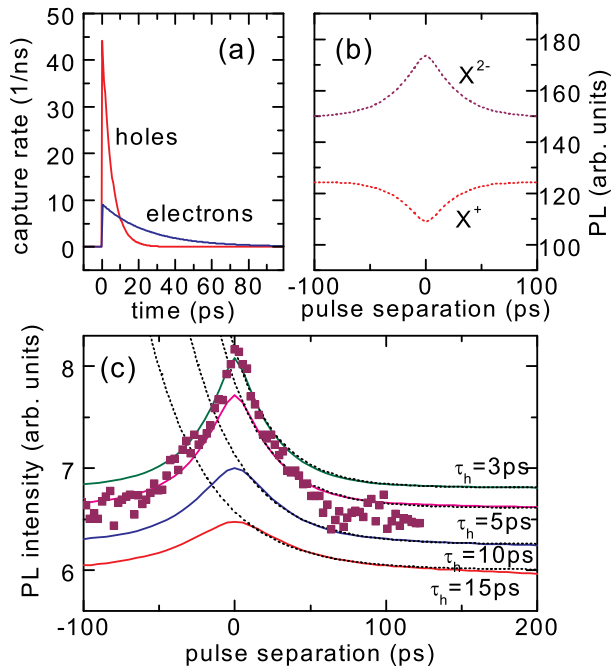}
\caption{(color online) Model of time dependence: (a) Capture rates for
both carrier types after a~single laser pulse. (b) Calculated
photoluminescence of X$^{2-}$ and X$^+$ lines for various delays in
two-pulse experiment. 
(c) X$^{2-}$ simulations for $\tau_e=25$ ps and different parameters
$\tau_h$. The curves are compared with experimental data (symbols) and
simple exponential relaxation with time-scale $\tau_e + \tau_h$.
\label{fig11}}
\end{figure}

In our continuous model, we consider 10 states of the quantum dot
including ground and first excited states for the total QD charge +1, 0,
-1, -2, and second excited states for the charge 0 and -1. This selection
is based on the identification of the optical transitions observed in the
spectrum. Following Ref. \onlinecite{suff-prb_korelacje},
we assume excitation by trapping of an~electron, a~hole or an~entire
exciton. Relative integrated rates of these processes are taken from Ref.
\onlinecite{suff-prb_korelacje}. Due to the double pulse excitation, the
temporal profile of excitation of each type is a~sum of two exponential
decays starting at arrival of subsequent laser pulses. The rate equations
that include thus defined excitation and radiative recombination are
integrated numerically. 
Photoluminescence intensities of various lines are found by integrating
the respective radiative recombination over one repetition period, after
finding a steady state of the system.

Example temporal profiles, calculated for the extreme charge states (X$^+$
and X$^{2-}$), are presented in Figure \ref{fig11}(b). A sharp increase of
the intensity of X$^{2-}$ is accompanied by a decrease of the X$^+$
intensity. Both features have a similar shape which can be approximated by
an exponential function. The characteristic time of this function is
determined by parameters $\tau_h$ and $\tau_e$. However it is never smaller
than the larger of them and it is close to the value of their sum : $\tau_h
+ \tau_e$.
The amplitude of the peak decreases when values $\tau_h$ and $\tau_e$
become closer to each other. A comparison of the characteristic times
obtained from simulations to the experimental ones allows us to conclude
that the electron capture time (the longer one) is in the range of 20-40ps.
The hole capture time is much smaller to assure a sufficient amplitude of
the observed features.

\section{Summary}

We performed a detailed time-resolved spectroscopic study of single
CdTe/ZnTe quantum dots. The excitation dynamics was investigated by the
time resolved  micro-photoluminescence, single photon correlation and
subpicosecond excitation correlation measurements. The time resolved
experiments were done with excitation by ultrafast pulsed laser working at
energy above the energy gap of the barrier material.
The obtained temporal profiles of excitation correlation exhibit several
characteristic features: sub-nanosecond decrease of the intensity common
for most of the PL lines (with characteristic time comparable to PL decay
time), and picosecond variation of the relative intensity of the lines
related to excitons of different charge state.
We propose a model describing observed features and demonstrate that it
requires that the carriers are trapped separately. Moreover the capture of
carriers of different charge take place at different delay from excitation.
The detailed analysis of the temporal profiles let us conclude that the
electron capture takes place in 20-40ps after excitation, while capture of
hole is much faster.

\begin{acknowledgments}
This work was partially supported by the Polish Ministry of Science and
Higher Education as research grants in years 2006-2009 and by European
Project No. MTKD-CT-2005-029671. One of us (PK) acknowledges the
support from European Project No. FP7/2007-2013-221515 (MOCNA).

\end{acknowledgments}

\end{document}